\documentclass[11.5pt]{article}
\usepackage[utf8]{inputenc}
\usepackage[T1]{fontenc}
\usepackage{multirow}
\usepackage{fullpage}
\usepackage{subfigure}
\usepackage{amsthm}
\usepackage{amssymb}
\usepackage{latexsym}
\usepackage{amsmath}
\usepackage{color}
\usepackage{cases}
\usepackage{graphicx}
\usepackage{setspace}
\usepackage{float}
\usepackage{natbib}
\usepackage[unicode=true, bookmarks=true, bookmarksnumbered=false, bookmarksopen=false, breaklinks=true, pdfborder={0 0 0},backref=false]{hyperref}
\hypersetup{
  colorlinks=true,
  allcolors=blue
}

\def\R{\mathbb{R}}

\begin{document}

\title{Conflicts, Assortative Matching, and the Evolution of Signaling Norms
\thanks{We sincerely thank two anonymous referees for their constructive suggestions that greatly improve the paper. We also thank Wallice Ao, Jonathan Newton, Van Kolpin and Anne van den Nouweland for their comments.
}}

\author{Ethan Holdahl\thanks{Department of Economics, University of Oregon, eholdahl@uoregon.edu} \and Jiabin Wu\thanks{Department of Economics, University of Oregon, jwu5@uoregon.edu}}
\date{March 20, 2023}
\maketitle

\begin{abstract}


This paper proposes a model to explain the potential role of inter-group conflicts in determining the rise and fall of signaling norms. Individuals in a population are characterized by high and low productivity types and they are matched in pairs to form social relationships such as mating or foraging relationships. In each relationship, an individual's payoff is increasing in its own type and its partner's type. Hence, the payoff structure of a relationship does not resemble a dilemma situation. Assume that types are not observable. In one population, assortative matching according to types is sustained by signaling. In the other population, individuals do not signal and they are randomly matched. Types evolve within each population. At the same time, the two populations may engage in conflicts. Due to assortative matching, high types grow faster in the population with signaling, yet they bear the cost of signaling, which lowers their population's fitness in the long run. Through simulations, we show that the survival of the signaling population depends crucially on the timing and the efficiency of weapon used in inter-group conflicts.
\end{abstract}
\newpage
\section{Introduction}
Harmful and wasteful practices such as elaborated body tattooing and piercings, lethal initiation rituals, excessive feasts, and wearing obstructive dressing codes have been observed throughout the human history. Why did humans adopt such practices? The classic theory of signaling by \cite{Zahavi1975} in biology and \cite{Spence1973} in economics provides an answer:\footnote{The theory has been widely applied to explain different phenomena ranging from life sciences to social sciences. See further discussion of the subject in \cite{Grafen1990}, \cite{MaynardSmithHarper1995}, \cite{Johnstone1997},  \cite{ZahaviZahavi1997}, \cite{MaynardSmithHarper2003}, \cite{SearcyNowicki2005}, \cite{Getty2006}, \cite{Grose2011} and \cite{Szamado2012}, among many others.} They serve as costly signaling devices to differentiate between types of individuals. Recently, \cite{PrzepiorkaDiekmann2021} call these practices the signaling norms. 

A more subtle yet important question is, suppose no signaling norm is the ancestral condition, what determined the emergence of the signaling norms? On the one hand, adopting a signaling norm to assort its members allowed a society to organize its social hierarchy, mating and reproduction in a more efficient way, which might help boosting the society's population growth rate. On the other hand, the cost of practicing the signaling norm might slow the rate down. Larger societies often had a higher chance to survive in ancient warfare and violent conflicts are argued to have played a critical role in human's ancestral past \citep{Keeley1996, WranghamPeterson1996,  BussShackelford1997, LeBlancRegister2003, GuilaineXammit2004, Gat2006, PottsHayden2008, Ferguson2012}. Would conflicts matter for the selection of the signaling norms? If so, how? In this paper, we attempt to formally investigate the role of conflicts in influencing the evolution of signaling norms in an evolutionary game theoretical model. 

In the model we propose, there are two populations. In each population, there are two types of individuals: high and low. The types are genetically determined and the high type has some hidden fitness advantages over the low type. The individuals need to match in pairs to obtain payoffs (forming a mating or foraging pair for example), which in turn determine their reproduction rates. However, the individuals cannot directly observe others' types. Assume that one population is equipped with signaling technology, while the other is not. In the population without signaling technology, the individuals are randomly matched in pairs because they cannot assort according to types. In the population with signaling technology, the high type individuals may adopt the signal to differentiate themselves from the low type individuals so that they can identify one another and avoid being matched with the low type individuals. As long as doing so is beneficial to the high type individuals, and at the same time, the cost of signaling is sufficiently high to deter the low type individuals to mimic the high type individuals, a separating equilibrium can be sustained in the population. Note that we treat signaling as a behavioral trait instead of a genetic trait, so the individuals can choose whether to adopt it in the behavioral timescale. 

The assortative benefit to high types in the signaling population allows them to evolve faster than they would without signaling. However, even absent signaling, high types would eventually prevail given their fitness advantage. When both populations are homogeneously high type, the assortative advantage the signaling population had over the non-signaling population disappears leaving the signaling population with a cost but no comparative benefit. This means the signaling population's fitness is lower in the long run. 
Therefore, there is a trade-off in utilizing signaling and it becomes crucial if the two populations engage in conflicts. Through simulations, we show that the timing and the the efficiency of weapon used in inter-group conflicts play an essential role in determining the survival of the signaling population. In particular, the signaling population has a higher chance to survive when it still has a fitness advantage if the period of isolation before conflicts is shorter and weapon used in conflicts is more efficient. 

Our model provides a rationale for why signaling norms---even if transitory---may appear in human history: A signaling norm helps to accelerate the population growth rate initially, which may give an advantage to a population during conflicts. However, eventually when high types dominate, it becomes redundant and there is no point for people to continue using it.



The theory of signaling has been used in different evolutionary models to explain human behavior. \cite{Gintisetal2001} formalize the idea by \cite{Zahavi1975} and \cite{Miller2000} that high-quality individuals may engage in pro-social activities (cooperate in a public good provision game) to signal their desirability to gain better mating opportunities. This is called the ``competitive altruism" in \cite{Roberts1998}. While quality can be signaled by doing good, it can well be signaled by any costly activity. \cite{Hopkins2014AEJMICRO} instead considers a model in which the individuals do not differ in qualities, but in their degrees of altruism. In this model, it is natural for the altruists to use pro-social activities to signal their altruism and the author shows that altruists who can mentalize have a greater advantage over nonaltruists. \cite{PrzepiorkaDiekmann2021} consider a trust game in which there are two types of trustees: short-term (impatient) and long-term (patient). They identify the condition under which a separated equilibrium with only the long term trustees send a costly signal to the trusters and they show that only when the probability of meeting a long-term trustee is sufficiently low for the trusters, would the separated equilibrium be collectively more beneficial than the case without any signaling opportunity. 
Note that these papers all consider what is called the ``dilemma" situations. Without signaling opportunity, ``
bad" type or behavior such as nonaltruist and defection prevail. On the contrary, we consider a non-dilemma situation and high types would eventually dominate the entire population with or without signaling opportunity.  In addition, these papers do not explicitly consider the role of conflicts, while we investigate how different properties of conflicts can affect the evolution of signaling norms. Many social relationships do not necessarily resemble ``dilemma" situations and they are understudied because of perceived triviality. Nevertheless, we show that interesting phenomena can still arise in such relationships.

The paper is organized as follows. Section 2 introduces the model, analyzes how types evolve within a population and how conflicts work between the two populations. Section 3 conducts comparative statics on the timing and the efficiency of weapon used in inter-group conflicts. Section 4 discusses alternative modeling choices and provides some concluding remarks.

\section{The Model and Analysis}

In this section we outline how populations evolve in our model. We are interested in the effect that signaling has on the evolution of a population so we consider two populations: one with signaling technology and the other without signaling technology. Both populations are endowed with high type individuals who are better suited to their environment and thus have a higher fitness, and low type individuals who are not as adapted to the current environment and thus have a lower fitness. The types are genetically determined, meaning that the individuals cannot change their own types and the distribution of types in a population evolves through reproduction. We use a discrete generational model where in each generation (period) individuals match with another individual in their population. This matching can be interpreted as a mating relationship, a foraging relationship, or any other social relationship. The type of each individual is not observable to the other members of the population, so where there is no signaling, individuals match randomly with another individual in their population. However, when signaling is present individuals are able to discriminate. As a result, signalers randomly match with another signaling individual in their population leaving the non-signalers in that population to randomly match with each other. After matching, they reproduce according to the payoffs in the table below minus the cost of signaling if applicable. Table \ref{tab:payofftable} reports the payoffs. An individual of type $i$, who is matched with an individual of type $j$, obtains a payoff of $V(i, j)$, for $i, j \in \{H, L\}$.  
Note that since these payoffs determine the individuals' life long reproductive rates, the length of each discrete period can be thought of as the amount of time needed for a given individual to reach adulthood.

\begin{table}[!ht]
\caption{Payoffs: Reproduction Rates}
\label{tab:payofftable}
\centering
\vskip6pt
 \begin{tabular}{cc|c|c|} 
    
      & \multicolumn{1}{c}{} & \multicolumn{1}{c}{H}  & \multicolumn{1}{c}{L} \\\cline{3-4}
      \multirow{2}*{}  & H & $V(H,H)$ & $V(H,L)$ \\\cline{3-4}
      & L & $V(L,H)$ & $V(L,L)$ \\\cline{3-4}
    \end{tabular}
    \vskip6pt
\end{table}

We impose the following relation on the payoffs: 
\begin{equation}\label{payoffrelation1}
 V(H,H)>V(H,L)>V(L,H)>V(L,L).
 \end{equation}

Viewing the matching as a mating relationship, we have a model where there are no hybrids, only high or low type offspring can be produced. High types produce $V(H,*)$ high type offspring which depend on who they match with. Likewise, low types produce $V(L,*)$ low type offspring which depend on who they match with. Then two high types will together produce $2V(H,H)$ high types, a low and high type pairing will produce $V(H,L)$ high types and $V(L,H)$ low types, and a 2 low type pairing will produce $2V(L,L)$ low types.

Alternatively, we can think of $V(*,*)$ as the amount of offspring per parent that makes it to reproduction. Since high types are more fit than low types, high types will be more successful than low types in making it to adulthood thus explaining $V(H,L) > V(L,H)$, even if we expect high and low types to be born in equal proportion in a mixed parent situation. Additionally, we can think that the parents provide for their offspring in some capacity before they reach adulthood. It is likely that high type parents are better providers than low type parents which explains $V(H,H) > V(L, H)$, $V(H, H)>V(H, L)$, $V(H, L)>V(L, L)$, and $V(L,H) > V(L,L)$. In other words, one's reproductive rate is increasing both in one's type and in its matched partner's type. 

An analogous story can be told in a foraging relationship. Matched individuals share some, but not all of their foraged goods together. Here, high types are more successful than low types in procuring food. Since the amount of food is directly related to reproductive rate, Inequality (\ref{payoffrelation1}) follows. Here, high types produce only high type offspring and low types produce only low type offspring either asexually or otherwise.\footnote{Note that joint foraging is usually considered as a typical example of a game with a dilemma and yet Inequality (\ref{payoffrelation1}) does not reflect a dilemma situation. We want to emphasize that $H$ and $L$ are not strategies, but types of the individuals, and if we consider that different individuals are matched to play a game of dilemma, their equilibrium payoffs as functions of their types would match Inequality (\ref{payoffrelation1}). Suppose when two individuals are matched, they play a prisoner's dilemma type foraging game with two strategies: exerting a high effort or exerting a low effort. The low effort is the strictly dominant strategy in the game. Hence, both individuals in a pair always choose to exert the low effort and $V(x,y)$ is the equilibrium payoff of an $x$-type individual against a $y$-type individual when both exert the low effort, where $x\in\{H,L\}$. Assume that the $H$-type individual exerting the low effort is still more productive than the $L$-type individual does and the two individuals in a pair are not sharing food equally but according to their productivity. Then we would still have Inequality (\ref{payoffrelation1}).}

In addition, we want to ensure that populations will not die out. Hence, we require $V(H,H)>1$.

\subsection{A Population without Signaling}

In a population without signaling technology, the  individuals cannot observe others' types. As such they are randomly matched with another individual in their population. We define the non-signaling population at time $t$ as $N_t \in \R$ where $N_t = N^H_t + N^L_t$. Here, $N^H_t \in \R$ and $N^L_t \in \R$ are the amount of high types and low types, respectively, in the non-signaling population at time $t$.

Assuming a large $N_t$, the law of large numbers implies that $N^H_t$ and $N^L_t$ evolve according to their expected payoffs from Table \ref{tab:payofftable}. So we have: 
\begin{eqnarray}
    && N^H_{t+1}=[\frac{N^H_t}{N_t}*V(H,H)+\frac{N^L_t}{N_t}*V(H,L)]*N^H_t, \label{nosignaldynamic1}\\
    && N^L_{t+1}=[\frac{N^H_t}{N_t}*V(L,H)+\frac{N^L_t}{N_t}*V(L,L)]*N^L_t.\label{nosignaldynamic2}
\end{eqnarray}

\begin{figure}[p]
  \caption{Population Dynamics Without Signaling}
   \label{fig:No_Signal}
    \includegraphics[width=\textwidth, height=.28\textheight]{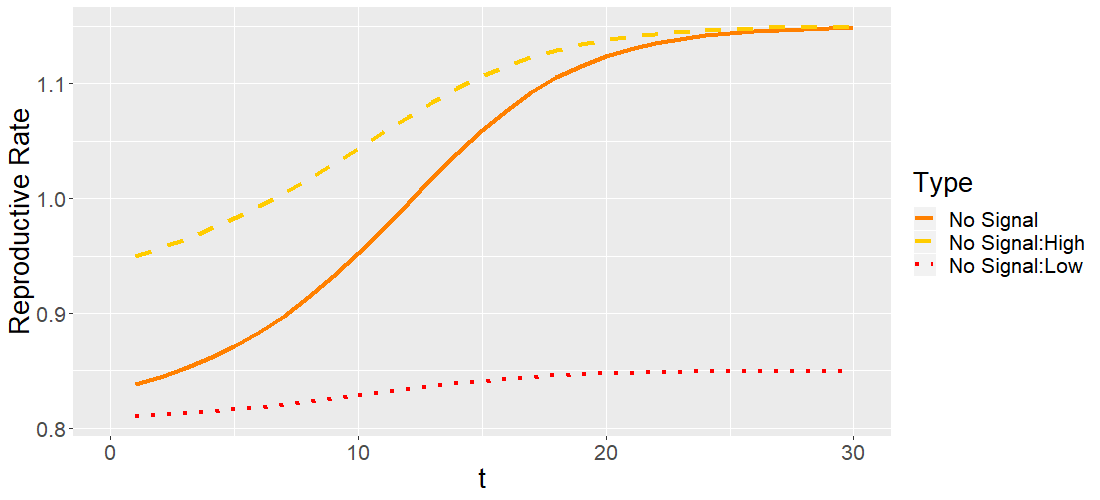}
    \includegraphics[width=\textwidth, height=.28\textheight]{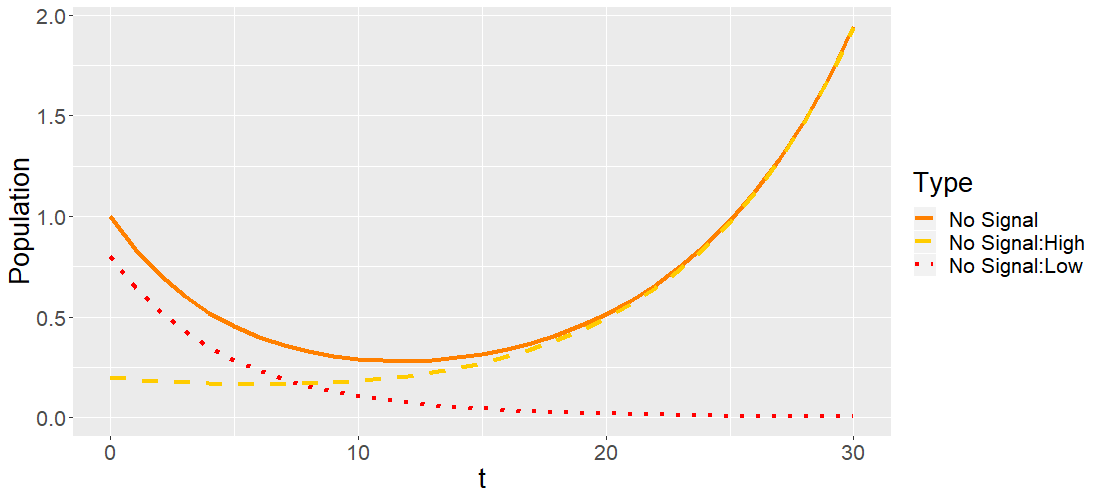}
    \includegraphics[width=\textwidth, height=.28\textheight]{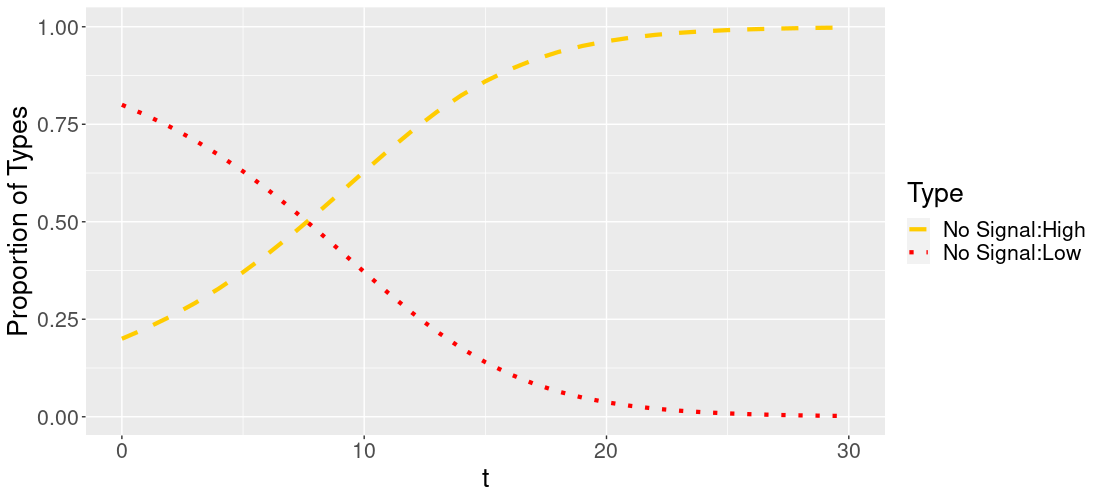}
    \begin{minipage}[c]{.2\textwidth}
    \textbf{Parameters:}
    \end{minipage}\hfill
    \begin{minipage}[c]{.2\textwidth}
    $N_0^H = .2$
    
    $N_0^L = .8$
    \end{minipage}\hfill
  \begin{minipage}[c]{.6\textwidth}
  \begin{tabular}{cc|c|c|}
      & \multicolumn{1}{c}{} & \multicolumn{1}{c}{H}  & \multicolumn{1}{c}{L} \\\cline{3-4}
      \multirow{2}*{}  & H & $1.15$ & $0.90$ \\\cline{3-4}
      & L & $0.85$ & $0.80$ \\\cline{3-4}
    \end{tabular}
    \end{minipage}
    \end{figure}

We simulate the dynamics described by Equations (\ref{nosignaldynamic1}) and (\ref{nosignaldynamic2}) in Figure \ref{fig:No_Signal}. The first graph shows the evolution of the average reproductive rate for each type as well as for the population as a whole and the second shows the evolution of the sizes of high and low type sub-populations as well as the population as a whole. Note that the starting population has been normalized to 1 with high types initially making up 20\% of the population.The second graph shows how the population level, and the amount of each type evolves over time. The final graph shows how the proportion of types within the population change over time.

Inequality (\ref{payoffrelation1}) implies $V(H,H)>V(L,H)$ and $V(H,L)>V(L,L)$. Hence, in the absence of signaling, high types are able to evolve within the population because they realize higher payoffs than low types i.e. survival of the fittest. As the proportion of high types increases in the population, the average reproductive rate of the population will converge to the reproductive rate of the high types. Note that there is a secondary effect here. Inequality (\ref{payoffrelation1}) also implies $V(H,H)>V(H,L)$ and $V(L,H)>V(L,L)$. Hence, as high types make up a greater proportion of the population, individuals become more likely to match with high types, thus the reproductive rates of both low types and high types increase as high types come to make up a greater proportion of the population, which can be seen in the first graph of Figure (\ref{fig:No_Signal}). 

\subsection{A Population with Signaling}

 In this section we consider a population that has signaling technology. We should emphasize that here we examine behavioral signaling that individuals can choose whether to opt into. Hence, the individuals make their decisions on signaling within each generation, while the genetically determined types evolve across generations. Let $K$ be the cost of signaling. 
 We assume that the cost of signaling, $K$, is sufficiently high such that only high types can afford it.  Formally, we require: \begin{equation} \label{IC1}
     V(L,L)>V(L,H)-K.
 \end{equation}
 
Inequality (\ref{IC1}) states that the payoff for a low type by being matched with another low type is larger than the benefit of adopting the signal, a higher payoff by being matched with a high type, minus the cost of signaling. We also only want to consider signals that are potentially incentive compatible for high types. That is, a high type receives a higher payoff by being matched with another high type, even after paying the cost of signaling, than it does by being matched with a low type: \begin{equation}\label{IC2}
     V(H,H)-K>V(H,L).
 \end{equation}
 
 Combining the Inequalities \ref{IC1} and \ref{IC2}, we know that for a viable signal of cost $K$ to exist it must be the case that: \begin{equation}
     V(H,H)+V(L,L)>V(L,H)+V(H,L).
 \end{equation}
 This is known as the single crossing property \citep{Zahavi1975, Spence1973}.
 We enforce these conditions on the reproductive rates $V(.,.)$ and the cost of signaling $K$.
 
 Since we consider behavioral signaling, we assume that the individuals reach an equilibrium on their decisions on adopting the signal within each generation before reproduction. In an equilibrium, each individual has no incentive to deviate from its current choice of whether to signal. As shown in \cite{Spence1973}, given the incentive compatibility conditions (\ref{IC1}) and (\ref{IC2}), there exists a separating equilibrium in which only high types signal. Moreover, all high types are matched with one another, and low types are matched with one another, resulting in perfectly assortative matching in types. The high types earn a payoff of $V(H, H)-K$ and the low types earn a payoff of $V(L, L)$ in equilibrium. Note that the separating equilibrium is independent of the distribution of types in the population. Even when the group of low types is vanishingly small, the high types would still pay the cost of signaling to segregate themselves from the low types.

 We define the population with signaling at time $t$ as $S_t \in \R$ where $S_t = S^H_t + S^L_t$. Here, $S^H_t \in \R$ and $S^L_t \in \R$ are the amount of high types and low types, respectively, in the signaling population at time $t$. Because of signaling, the individuals are only matched with their own types. Hence, $S^H_t$ and $S^L_t$ experience a constant reproductive rate over time. Their evolutionary dynamics are as follows: 
\begin{eqnarray}
   && S^H_{t+1}=[V(H,H)-K]*S^H_t, \label{signaldynamic1}\\
&&     S^L_{t+1}=V(L,L)*S^L_t. \label{signaldynamic2}
\end{eqnarray}

\begin{figure}[p]
  \caption{Population Dynamics Where High Types Signal}
   \label{fig:Signal}
    \includegraphics[width=\textwidth, height=.28\textheight]{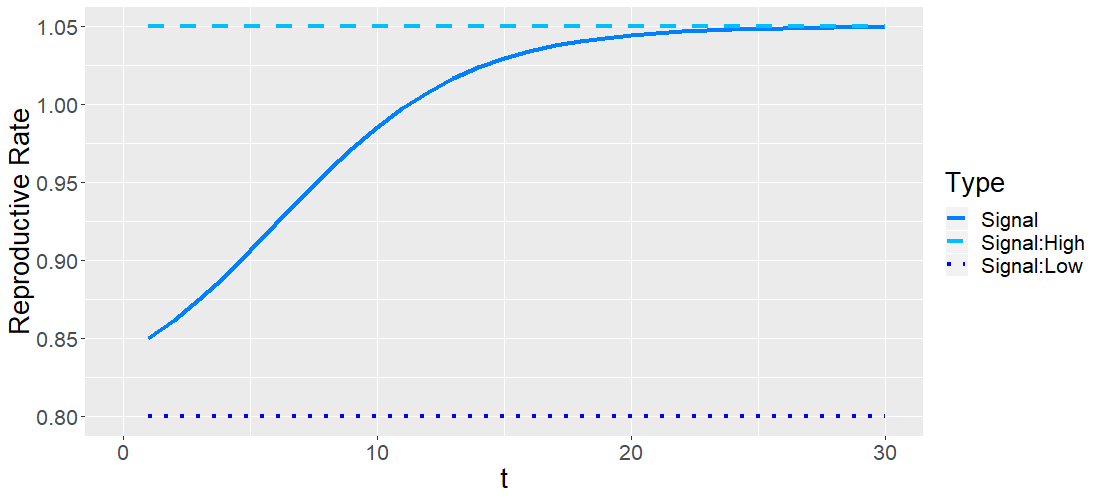}
    \includegraphics[width=\textwidth, height=.28\textheight]{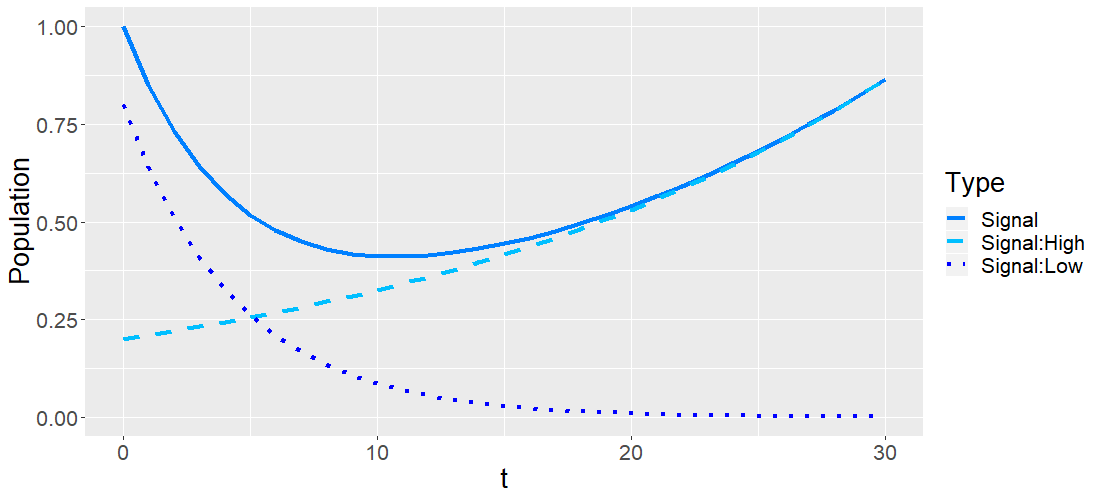}
    \includegraphics[width=\textwidth, height=.28\textheight]{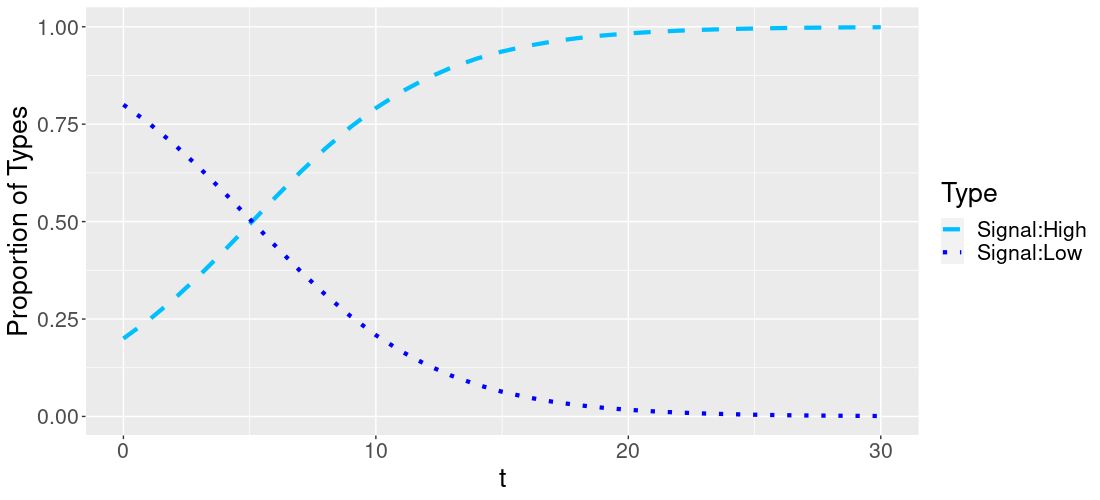}
 \begin{minipage}[c]{.2\textwidth}
    \textbf{Parameters:}
    \end{minipage}\hfill
    \begin{minipage}[c]{.2\textwidth}
    
    $S_0^H = .2$
    
    $S_0^L = .8$
    
    $K = .1$
    \end{minipage}\hfill
  \begin{minipage}[c]{.6\textwidth}
  \begin{tabular}{cc|c|c|}
      & \multicolumn{1}{c}{} & \multicolumn{1}{c}{H}  & \multicolumn{1}{c}{L} \\\cline{3-4}
      \multirow{2}*{}  & H & $1.15$ & $0.90$ \\\cline{3-4}
      & L & $0.85$ & $0.80$ \\\cline{3-4}
    \end{tabular}
    \end{minipage}
    \end{figure}

We simulate the dynamics described by Equations (\ref{signaldynamic1}) and \ref{signaldynamic2} in Figure \ref{fig:Signal}. We have three similar graphs as in Figure (\ref{fig:No_Signal}) and use the same parameters. Note that in the top graph, the reproductive rate of the high types and low types are not influenced by the relative size of each sub-population since they always match with their own types. This can be seen in Equations (\ref{signaldynamic1}) and (\ref{signaldynamic2}),  where the coefficients on $S^H_t$ and $S^L_t$, do not depend on the population composition. As a result, the reproductive rate of the population increases only because high types evolve to make up a greater proportion of the population. It can not be seen in the graphs here, but it should be clear that as $K$, the cost of signaling, decreases the speed of evolution increases.





\subsection{Comparing Signaling and Non-Signaling Populations}

\begin{figure}[p]
  \caption{Comparison of Population Dynamics Between Signaling and Non-Signaling Populations}
   \label{fig:Apart}
    \includegraphics[width=\textwidth, height=.28\textheight]{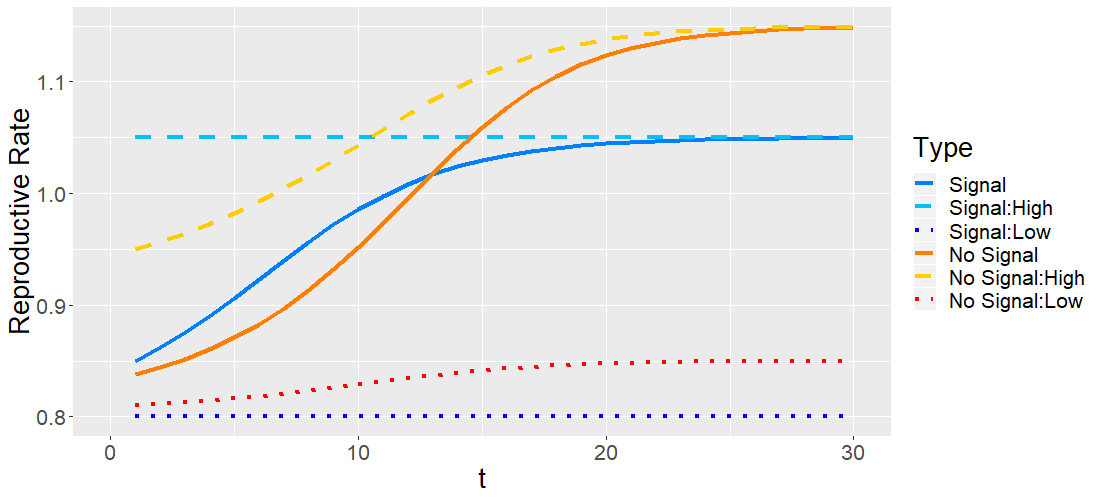}
    \includegraphics[width=\textwidth, height=.28\textheight]{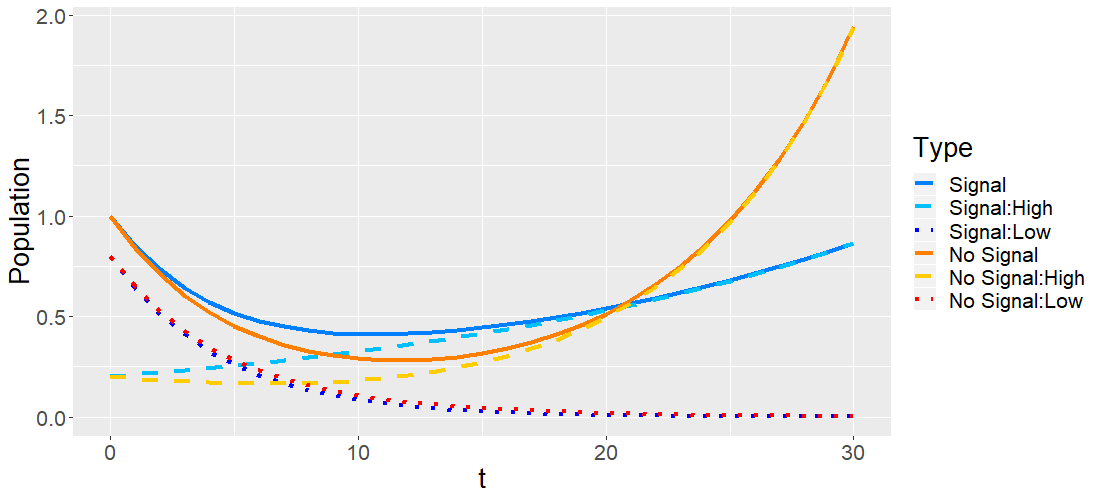}
    \includegraphics[width=\textwidth, height=.28\textheight]{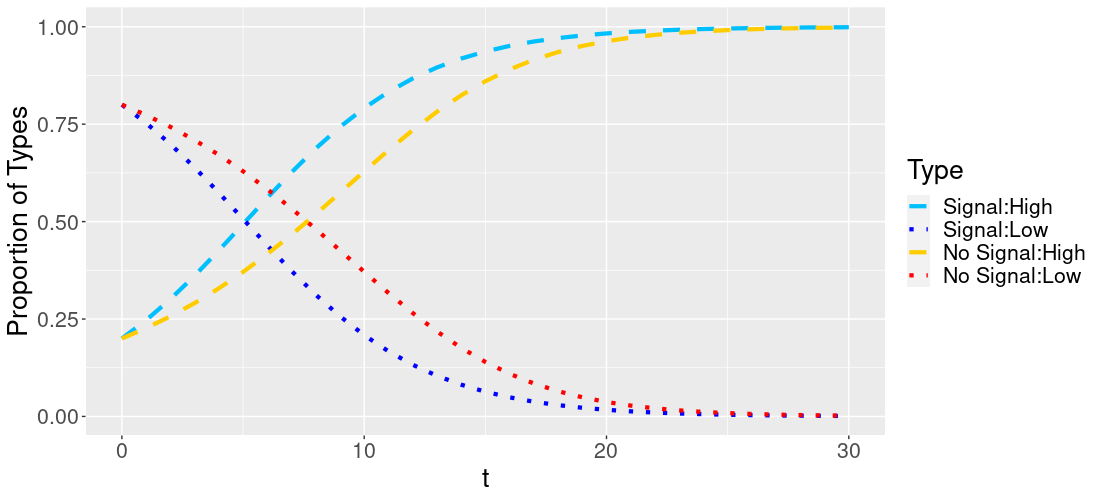}
 \begin{minipage}[c]{.2\textwidth}
    \textbf{Parameters:}
    \end{minipage}\hfill
    \begin{minipage}[c]{.2\textwidth}
    
    $N_0^H = S_0^H = .2$
    
    $N_0^L = S_0^L = .8$
    
    $K = .1$
    \end{minipage}\hfill
  \begin{minipage}[c]{.6\textwidth}
  \begin{tabular}{cc|c|c|}
      & \multicolumn{1}{c}{} & \multicolumn{1}{c}{H}  & \multicolumn{1}{c}{L} \\\cline{3-4}
      \multirow{2}*{}  & H & $1.15$ & $0.90$ \\\cline{3-4}
      & L & $0.85$ & $0.80$ \\\cline{3-4}
    \end{tabular}
    \end{minipage}
    \end{figure}

Figure \ref{fig:Apart} shows the overlap of Figure \ref{fig:No_Signal} and Figure \ref{fig:Signal}, both of which have the same parameters. Looking at the first graph of Figure \ref{fig:Apart}, we can see that while $N^H_t/N_t$ is sufficiently small, the reproductive rate for the signaling high types is greater than the reproductive rate for the high types in the non-signaling population. This is because $V(H,H) - K > V(H,L)$. As a result, high types evolve faster within the signaling population compared to the high types in the non-signaling population. In the first graph of Figure \ref{fig:Apart}, we can see that it takes until period 10 before the high types in the non-signaling population reach the same reproductive rate of their counterparts in the signaling population. Because of this and because low types in the signaling population have a lower reproductive rate than low types in the non-signaling population, high types make up a larger proportion of the signaling population compared to the non-signaling population. Therefore, it takes a few more periods until the average reproductive rate across the non-signaling population surpasses that of the signaling population. Looking at the second graph in Figure \ref{fig:Apart}, we can see that although the reproductive rate of the non-signaling population surpasses that of the signaling population around period 13, the signaling population retains a population advantage until around period 21. This is because the signaling population builds a significant population advantage in the initial periods. However, once sufficient time has passed and both populations are essentially homogeneous with only high types left, the non-signaling population realizes a reproductive rate that is $K$ greater than that of the signaling population. The final graph in Figure \ref{fig:Apart} shows the difference in the type makeup of the two populations. Note that the greatest difference occurs shortly after the start of the evolution but then vanishes as high types eventually make up the entirety of both populations.

\subsection{Conflicts between the Two Populations}

Assume that in period $T$ the non-signaling population and the signaling population engage in conflicts. In periods $t \in [0,T)$ the populations evolve according to the dynamics in the above sections, but starting at period $T$ they start to eliminate the other population according to the Lanchester's square law \citep{Lanchester1916}. The Lanchester's square law has been applied to a variety of human and non-human conflicts. See for example \cite{Wilsonetal2002}, \cite{JohnsonMackay2015EHB} and \cite{clifton2020brief}.\footnote{Another common way of modeling intergroup conflict is to treat it as a game played between two groups. Individuals within a group can contribute to the group's effort, which is costly to themselves, but beneficial to the group collectively in the conflict. Hence, an individual may have an incentive to free ride on fellow group members. See \cite{Bornstein2003} for a review. Since our focus is on signaling behavior, we refrain from complicating the model by adding an extra layer of effort choice.} In each period, before reproduction, each member of a population kills $\beta$ members of the other population. Individuals that are killed are drawn uniformly from the population. Parameter $\beta$ reflects how efficient the weapons used in the conflicts are. Hence, the dynamics for $N^H, N^L, S^H,$ and $S^L$ are now given as:
\begin{eqnarray}
    && N^H_{t+1}=[\frac{N^H_t}{N_t}*V(H,H)+\frac{N^L_t}{N_t}*V(H,L)]*max\{[N^H_t-\beta \frac{N^H_t}{N_t} I_{t \geq T}S_t],0\},
\\
&& N^L_{t+1}=[\frac{N^H_t}{N_t}*V(L,H)+\frac{N^L_t}{N_t}*V(L,L)]*max\{[N^L_t-\beta \frac{N^L_t}{N_t} I_{t \geq T}S_t],0\},
\\
    && S^H_{t+1}=[V(H,H)-K]*max\{[S^H_t-\beta \frac{S^H_t}{S_t} I_{t \geq T}N_t],0\},
\\
  && S^L_{t+1}=V(L,L)*max\{[S^L_t-\beta \frac{S^L_t}{S_t} I_{t \geq T}N_t],0\}.
\end{eqnarray}

In the equations above, $I_{t \geq T}$ is an indicator that the groups are engaging in inter-group conflicts and the max argument simply ensures that the populations would not reach a negative number. Note that we specify that conflicts take place before reproduction in each period in this model. It is clear that when applying this model, the time between each discrete period would simply be the amount of time it takes for a generation to reproduce. While conflicts may run at a different time scale, we can force conflicts into the reproduction time scale by adjusting the value of $\beta$. 

\begin{figure}[p]
  \caption{Conflicts Between Signaling and Non-Signaling Populations After a Period of Peace}
   \label{fig:Fight}
    \includegraphics[width=\textwidth, height=.28\textheight]{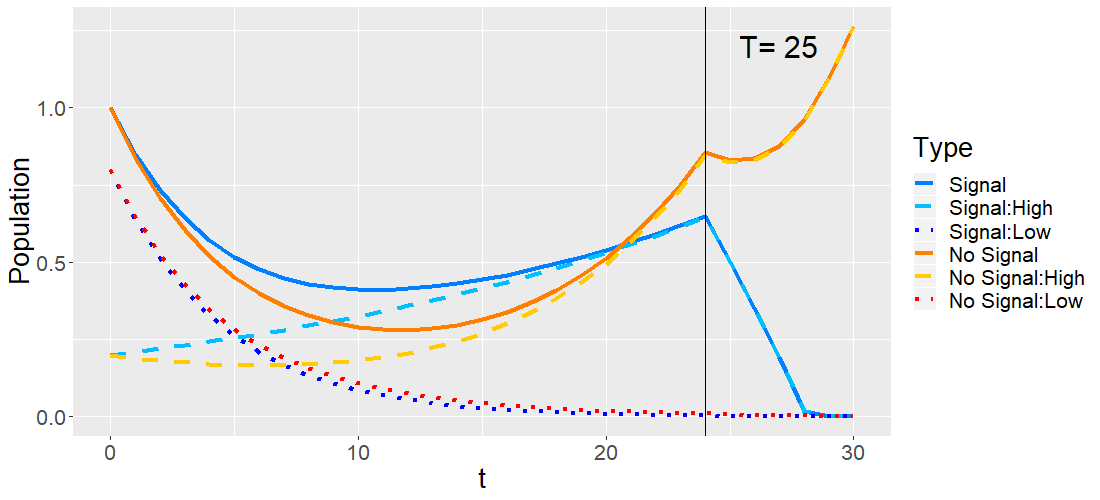}
    \includegraphics[width=\textwidth, height=.28\textheight]{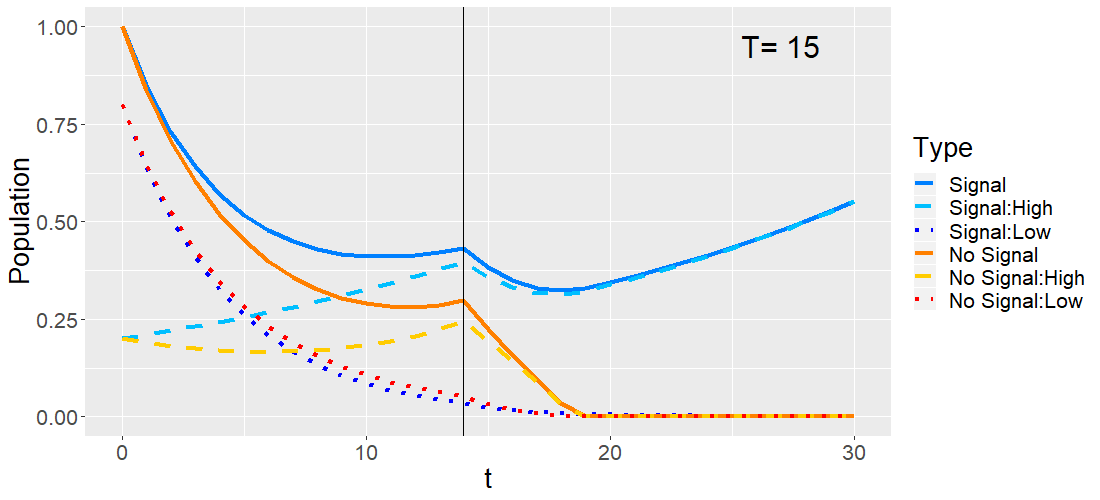}
    \includegraphics[width=\textwidth, height=.28\textheight]{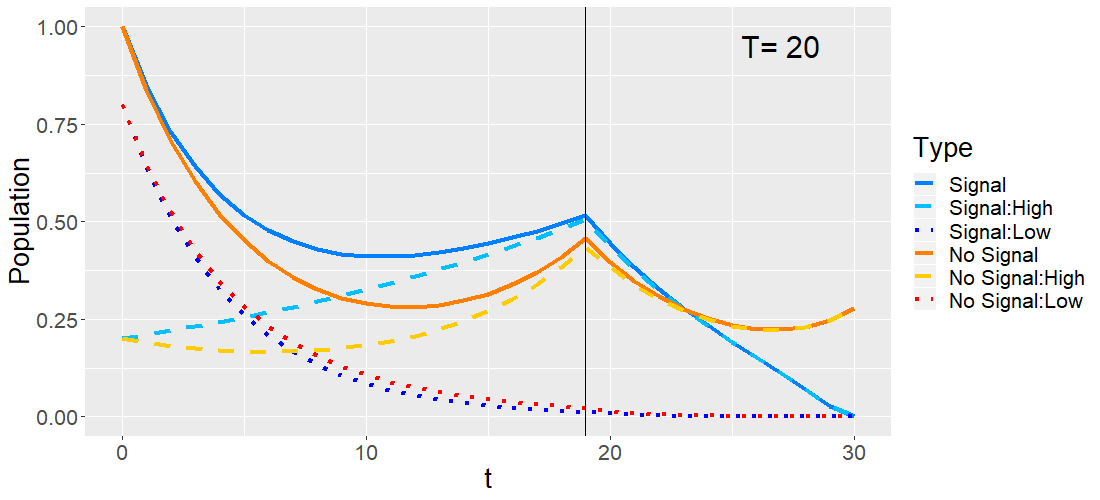}
 \begin{minipage}[c]{.2\textwidth}
    \textbf{Parameters:}
    \end{minipage}\hfill
    \begin{minipage}[c]{.2\textwidth}
    
    $N_0^H = S_0^H = .2$
    
    $N_0^L = S_0^L = .8$
    
    $K = .1$
    
    $\beta = .2$
    \end{minipage}\hfill
  \begin{minipage}[c]{.3\textwidth}
  \begin{tabular}{cc|c|c|}
      & \multicolumn{1}{c}{} & \multicolumn{1}{c}{H}  & \multicolumn{1}{c}{L} \\\cline{3-4}
      \multirow{2}*{}  & H & $1.15$ & $0.90$ \\\cline{3-4}
      & L & $0.85$ & $0.80$ \\\cline{3-4}
    \end{tabular}
    \end{minipage}\hfill
    \begin{minipage}[c]{.3\textwidth}
    The vertical lines indicate the last period before conflicts begin $(T-1)$.
    \end{minipage}
    \end{figure}

Figure \ref{fig:Fight} provides three examples of outcomes when conflicts occur in our model. In the top graph, the non-signaling population enters the periods of conflicts with a population advantage. Because they do not need to pay the cost of signaling, the fact that they have a population advantage indicates that they also have a superior reproductive rate. As a result, the non-signaling population always wins. In the second graph, the signaling population enters the conflicts with a population advantage. At this point the signaling population doesn't necessarily have a higher reproductive rate, however, if weapon used in conflicts is sufficiently efficient ($\beta$ is sufficiently large), the signaling population will be able to eliminate the non-signaling population before the non-signaling population can catch up with their superior long run reproductive rate. The third graph is a special case. Although the signaling population enters the conflicts with a slightly larger population, because $\beta$ is sufficiently small, the non-signaling population is able to overcome the signaling population during the conflicts.

\section{Comparative Statics}

\begin{figure}[p]
  \caption{Conflicts Between Signaling and Non-Signaling Populations After a Period of Peace: Comparative Statics}
   \label{fig:Regions}
    \includegraphics[width=\textwidth, height=.28\textheight]{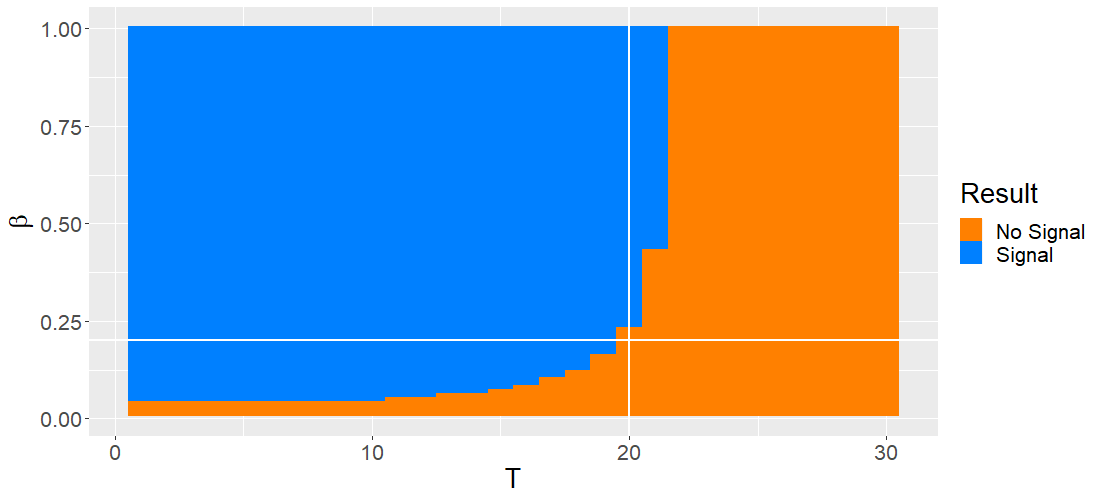}
    \includegraphics[width=\textwidth, height=.28\textheight]{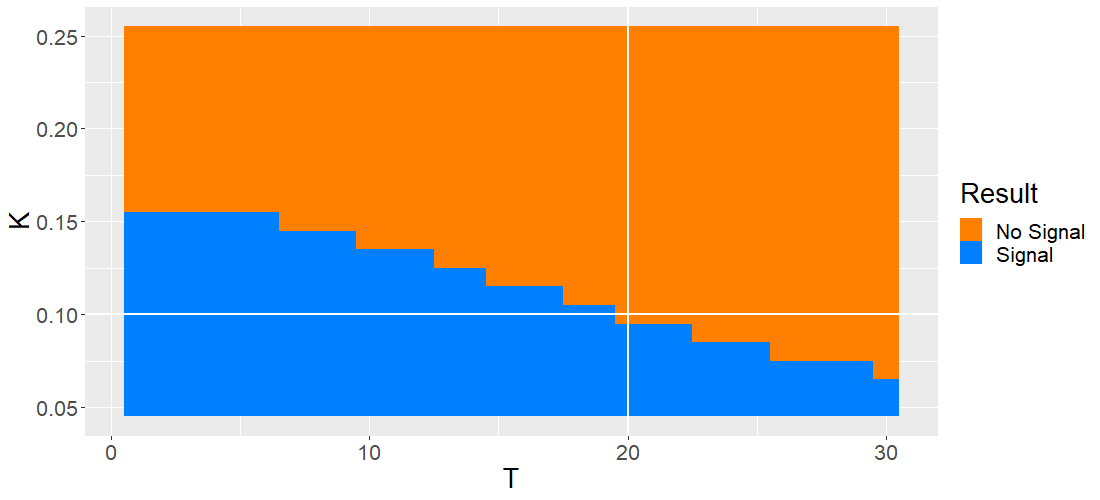}
    \includegraphics[width=\textwidth, height=.28\textheight]{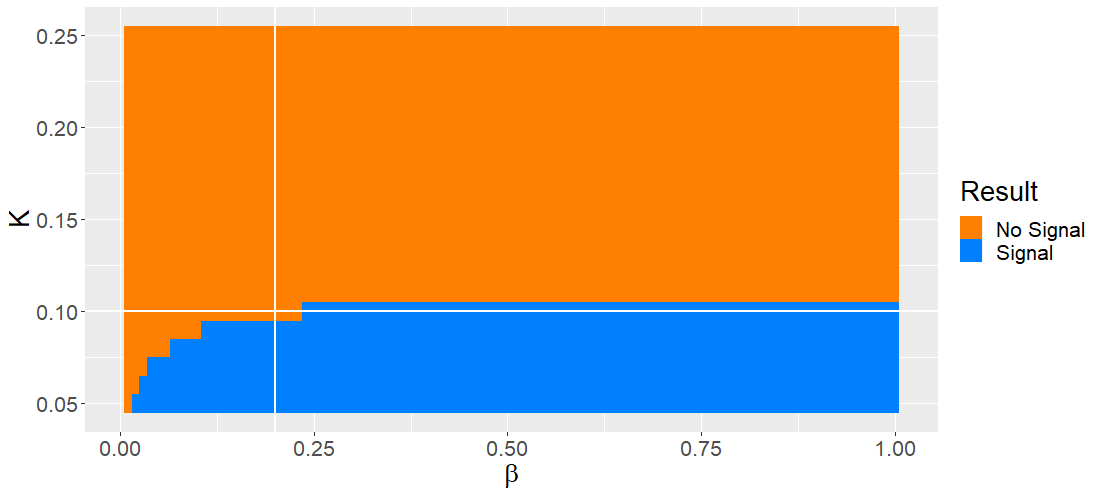}
 \begin{minipage}[c]{.2\textwidth}
    \textbf{Parameters:}
    \end{minipage}\hfill
    \begin{minipage}[c]{.2\textwidth}
    
    $N_0^H = S_0^H = .2$
    
    $N_0^L = S_0^L = .8$
    
    $K = .1$
    
    $\beta = .2$
    
    $T = 20$
    \end{minipage}\hfill
  \begin{minipage}[c]{.3\textwidth}
  \begin{tabular}{cc|c|c|}
      & \multicolumn{1}{c}{} & \multicolumn{1}{c}{H}  & \multicolumn{1}{c}{L} \\\cline{3-4}
      \multirow{2}*{}  & H & $1.15$ & $0.90$ \\\cline{3-4}
      & L & $0.85$ & $0.80$ \\\cline{3-4}
    \end{tabular}
    \end{minipage}\hfill
    \begin{minipage}[c]{.3\textwidth}
    The white intercept lines indicate the current parameters values. 
    \end{minipage}
    \end{figure}
    
We have shown that costly signaling in a population may be competitively advantageous in inter-group conflicts in our model. The viability of signaling depends largely on three different factors: the period that conflicts start ($T$), the cost of signaling ($K$), and the efficiency of weapon used in conflicts ($\beta$). Here, we argue that the signaling population benefits from a shorter period of isolation before conflicts, a smaller cost of signaling, and more efficient weapon used in conflicts. To clearly see the effects of these three factors, we run the model thousands of different times, varying the key parameters: $T,K$ and, $\beta$ and indicate the outcomes of the conflicts. The results are reported in Figure \ref{fig:Regions}, in which each pixel plotted corresponds to the outcome of the evolution given a set of parameters. In areas colored orange, the non-signaling population wins the conflicts and in the blue area the signaling population wins. The white intercept lines indicate where the model is evaluated in the last graph of Figure \ref{fig:Fight}.

The first graph in Figure \ref{fig:Regions} shows the result of the conflicts between the signaling population and the non-signaling population as we vary the period that conflicts start and the efficiency of weapon used in conflicts. As we can see, beyond a certain period the non-signaling population will always win no matter what the value of $\beta$ is. This is because the  non-signaling population is both greater and growing faster beyond that period. One can verify this by looking at Figure \ref{fig:Apart}. It is also necessary for $\beta$ to be sufficiently large for the signaling population to possibly win the conflicts. As we can see, when $\beta$ is sufficiently small, the non-signaling population always wins. Looking at the second graph of Figure \ref{fig:Regions}, we can see that a smaller signaling cost benefits the signaling population. The third graph of Figure \ref{fig:Regions} shows the interaction between the cost of signaling and the efficiency of weapon used in conflicts at $T=20$. The non-signaling population always wins when $K>.1$ because the non-signaling would have a larger population than the signaling population at $T=20$. Before that point, more efficient weapon used in conflicts can compensate for a higher cost of signaling for the signaling population.  

\begin{figure}[p]
  \caption{Conflicts Between Signaling and Non-Signaling Populations After a Period of Peace: Initial Conditions Comparative Statics}
   \label{fig:RatioRegions}
    \includegraphics[width=\textwidth, height=.28\textheight]{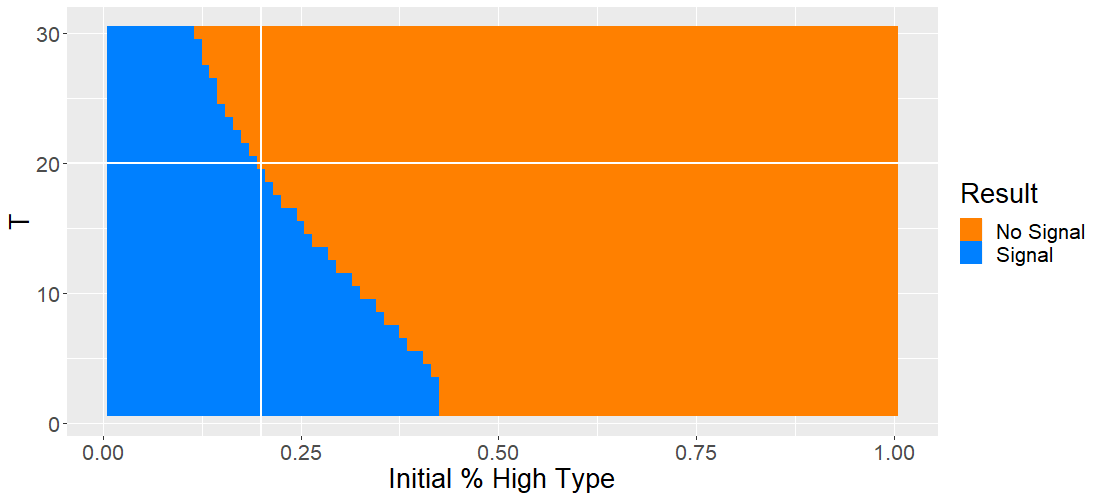}
    \includegraphics[width=\textwidth, height=.28\textheight]{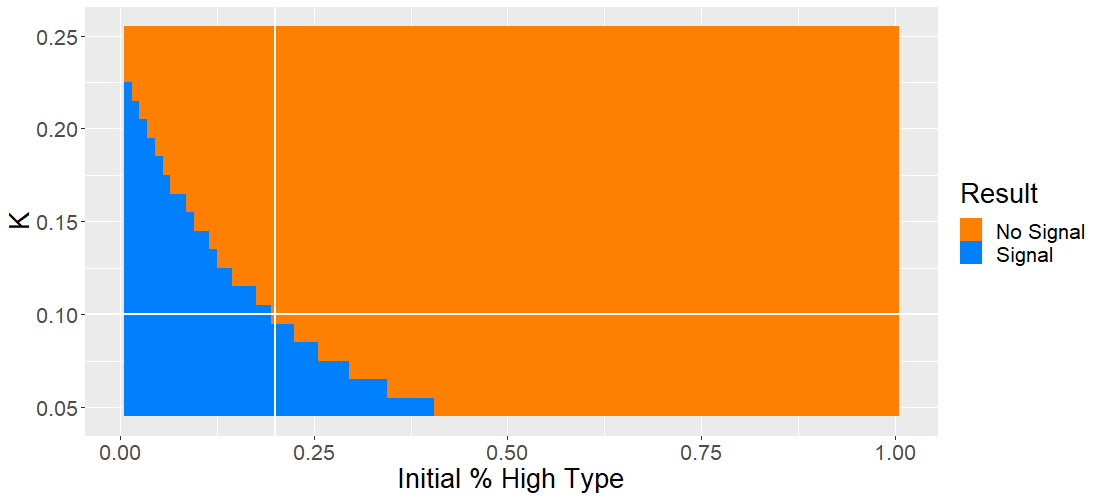}
    \includegraphics[width=\textwidth, height=.28\textheight]{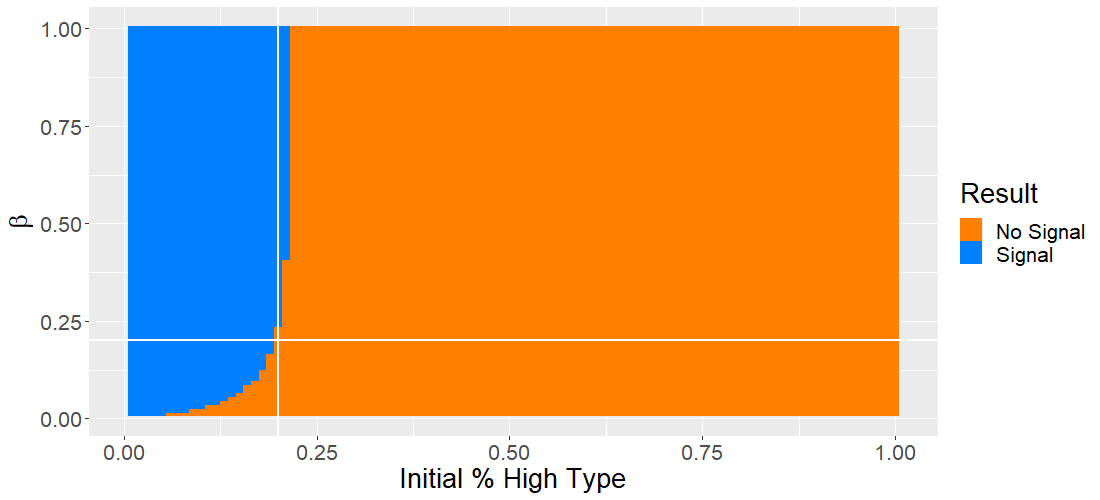}
 \begin{minipage}[c]{.2\textwidth}
    \textbf{Parameters:}
    \end{minipage}\hfill
    \begin{minipage}[c]{.2\textwidth}
    
    $N_0^H = S_0^H = .2$
    
    $N_0^L = S_0^L = .8$
    
    $K = .1$
    
    $\beta = .2$
    
    $T = 20$
    \end{minipage}\hfill
  \begin{minipage}[c]{.3\textwidth}
  \begin{tabular}{cc|c|c|}
      & \multicolumn{1}{c}{} & \multicolumn{1}{c}{H}  & \multicolumn{1}{c}{L} \\\cline{3-4}
      \multirow{2}*{}  & H & $1.15$ & $0.90$ \\\cline{3-4}
      & L & $0.85$ & $0.80$ \\\cline{3-4}
    \end{tabular}
    \end{minipage}\hfill
    \begin{minipage}[c]{.3\textwidth}
    The white intercept lines indicate the current parameters values. 
    \end{minipage}
    \end{figure}

Please note that thus far we have used $N_0^H = .2$ and $N_0^L = .8$ and $S_0^H = .2$ and $S_0^L = .8$. Recall that high types are the individuals that are better suited to their current environment and thus have a higher fitness than low types. With this in mind, it is useful to think about the initial values of $N_0^H, N_0^L, S_0^H,$ and $S_0^L$ as being related to the speed at which the environment is changing around the people. 
The faster the environment changes, the less time passes before a new genetic mutation is considered the high type which would essentially restart our model with a small proportion of high types. So, a faster changing environment would translate to a smaller initial proportion of high types. 
With that in mind, Figure \ref{fig:RatioRegions} shows the results of the conflicts when we vary the initial proportion of high types in each population, $N_0^H$ and $S_0^H$, while keeping them consistent across populations, $N_0^H = S_0^H$, and maintaining $N_0 = S_0 = 1$ so that $N_0^H$ and $S_0^H$ is simply the initial proportion of high types in each population. The results are clear: The smaller the initial proportion of high types, the more likely it is that the signaling population will win the conflicts. 
This is because the assortative advantage is largest when the proportion of high types is smallest which further exaggerates the speed at which the makeup of the population changes from low type to high type. Hence, giving the populations more room to evolve naturally favors the signaling population. Abstracting from the model, this implies that signaling is more advantageous the quicker one's environment changes. In a rapidly changing environment, the amount of high types at any given time is expected to be lower than in a slow changing environment since we define high types as those best adapted to the current environment. As can be seen in the graphs, the smaller the initial proportion of high types, the wider the range of $T, K$ and $\beta$ values that result in conflicts that favors the signaling population.

When competition erupts or is ongoing, the eventual winner can be determined by examining the relative population levels as well as the type distribution within those populations.

 \begin{figure}[h]
   \caption{Relative Population Needed for Signaling Population to Win in Conflict}
   \label{fig:FightProps}
    \includegraphics[width=\textwidth, height=.33\textheight]{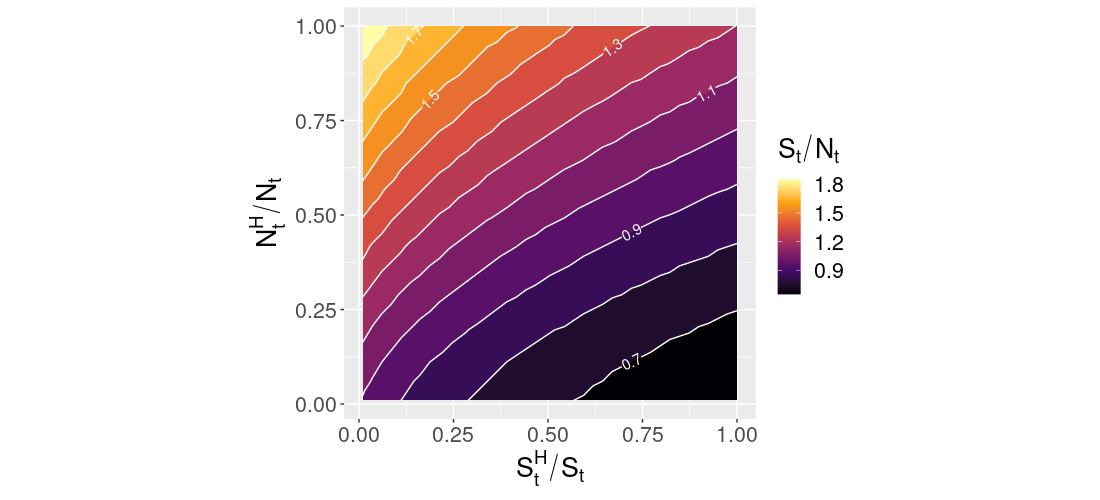}
    \begin{minipage}[c]{.2\textwidth}
    \textbf{Parameters:}
    \end{minipage}\hfill
    \begin{minipage}[c]{.2\textwidth}
    
    $K = .1$
    
    $\beta = .2$
    
    \end{minipage}\hfill
  \begin{minipage}[c]{.3\textwidth}
  \begin{tabular}{cc|c|c|}
      & \multicolumn{1}{c}{} & \multicolumn{1}{c}{H}  & \multicolumn{1}{c}{L} \\\cline{3-4}
      \multirow{2}*{}  & H & $1.15$ & $0.90$ \\\cline{3-4}
      & L & $0.85$ & $0.80$ \\\cline{3-4}
    \end{tabular}
    \end{minipage}\hfill
    \begin{minipage}[c]{.3\textwidth}
    \end{minipage}
\end{figure}

Figure \ref{fig:FightProps} depicts the minimum relative population level ($S_t/N_t$) needed for the signaling population to defeat the non-signaling population once conflict is ongoing given the type makeup of each population. Since a greater proportion of high types increases the reproductive rate in each population, as the proportion of high types in the signaling population, $S_t^H/S_t$ increases, a smaller relative population level, $S_t/N_t$ is needed for the signaling population to win. The inverse is true for increases in the proportion of high types in the non-signaling population, $N_t^H/N_t$.




\section{Discussions}

\subsection{Alternative Modeling Choices}
\subsubsection{Low Types also Signal}
In our model we assume that in the signaling population only high types transmit a signal and as a result, the signal enforces perfect assortative matching within the signaling population. One may question what if the signal is ubiquitous across the signaling population such that low types as well as high types transmit the signal. In this case, the signal would not be a signal at all but just a costly trait since no one would be able to distinguish one type from another. As a result, the one difference between the evolutionary dynamics of the signaling and non-signaling population would be that the signaling population experiences a level decrease in their reproductive rate with no functional assortativity advantage. The only scenarios where the signaling population is able the beat the non-signaling population in conflicts in this setup are with extreme parameter values where the signaling cost and the expression $V(H,H)-V(H,L)$ are so large that the low types in the signaling population are effectively eliminated by the signaling cost immediately, thus effectively giving the high types assortative matching which, if the initial proportion of high types in each population is low enough, may allow for a very narrow window under which the signaling population will realize a population advantage. 

\subsubsection{One Population of Signalers and Non-Signalers}
\begin{figure}[hp]
   \caption{One Population Where Some High Types Signal and Some Don't}
   \label{fig:One_Pop}
    \includegraphics[width=\textwidth, height=.28\textheight]{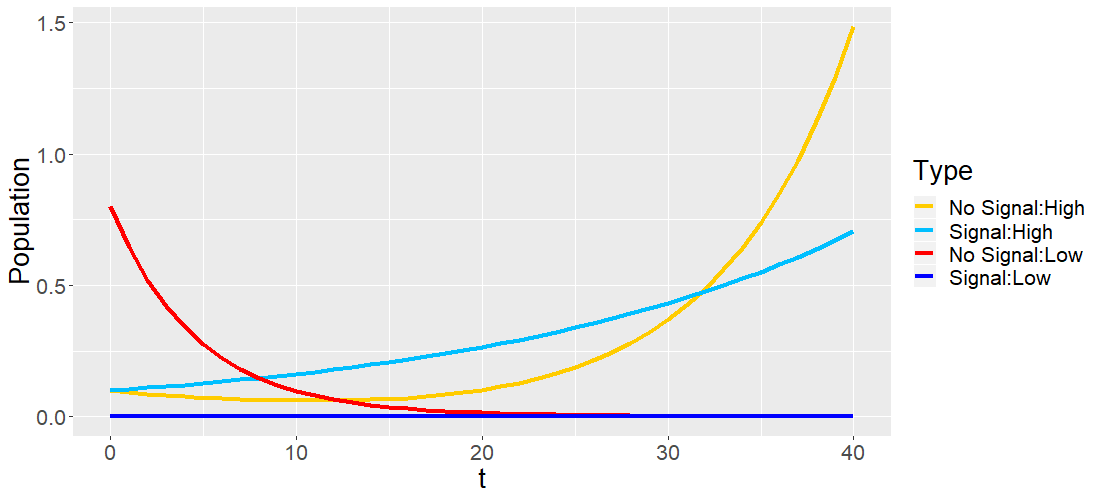}
 \begin{minipage}[c]{.2\textwidth}
    \textbf{Parameters:}
    \end{minipage}\hfill
    \begin{minipage}[c]{.2\textwidth}
    
    $N_0^H = S_0^H = .1$
    
    $N_0^L = .8$
    
    $S_0^L = 0$
    
    $K = .1$

    \end{minipage}\hfill
  \begin{minipage}[c]{.3\textwidth}
  \begin{tabular}{cc|c|c|}
      & \multicolumn{1}{c}{} & \multicolumn{1}{c}{H}  & \multicolumn{1}{c}{L} \\\cline{3-4}
      \multirow{2}*{}  & H & $1.15$ & $0.90$ \\\cline{3-4}
      & L & $0.85$ & $0.80$ \\\cline{3-4}
    \end{tabular}
    \end{minipage}\hfill
    \begin{minipage}[c]{.3\textwidth}

    \end{minipage}
    \end{figure}

Another question one may have is what if there are both signalers and non-signalers in the same population. In this case, so long as the signaling high types are able to gain an assortative advantage over the non-signaling high types in the population, then that may be enough to offset the signal cost and realize a higher reproductive rate for them in the short run even in the absence of inter-group conflicts. We demonstrate that this is not the case in a one-population model with four types (signaling high type, non-signaling high type, signaling low type, non-signaling low type) as shown in Figure \ref{fig:One_Pop}. The rationale is as follows. In our settings, high types evolve to make up an increasing proportion of the population with certainty as time increases until there are only high types remaining. As a result, at some point the relative difference in assortativity between the signaling high types and the non-signaling high types will not be big enough to offset the cost of signaling. Therefore, absent conflicts between the signalers and non-signalers, non-signaling high types will dominate the entire population with certainty in the long run.

The above discussions on alternative modelling choices suggest that signalling being behavioral (only high types choose to adopt it) and conflicts between the signaling population and the non-signaling population  are prerequisites for signaling to survive.

\newpage
\subsubsection{Imperfect Signaling}

In the dynamics explored in this paper, we assumed agents always made the payoff maximizing choice which led to perfect assortativity in the signaling population. This perfect assortativity allowed the signaling population to gain a temporary reproductive edge over the non-signaling population. Here we will relax the assumption of perfect associativity and instead examine how our model behaves where there is a probability $p$ that agents make a mistake and signal when they shouldn't (low types) or not signal when they should (high types). $1-p$ measures the accuracy of signaling.\footnote{Essentially, we endogenize the degree of assortative matching through imperfect signaling. See \cite{NaxRigos2016}, \cite{Newton2017IJGT}, \citet{wu2016,wu2018,wu2023wp} for the study of endogenous assortativity through different mechanisms.} Intuitively, $p$ should be some value in the range $[0,1/2)$ as $p=0$ implies mistake free signaling and $p=1/2$ implies signaling is entirely uninformative. As such, the payoffs realized by each type is no longer fixed and now depends on the proportion of high types to low types within the population as a whole. Note that the dynamics in the non-signaling population remain unchanged by this assumption. 

If high types imperfectly signal and low types imperfectly choose not to signal that means that $S^H_t*(1-p) + S^L_t*p$ describes the total number of signalers and $S^H_t*p + S^L_t*(1-p)$ describes the total number of non-signalers at time $t$.

Assuming that high types choose to signal and low types choose to not signal, and match according to their signal. When a high type does signal, the payoff they get is given by the following expression:
\setlength{\belowdisplayskip}{4pt} \setlength{\belowdisplayshortskip}{4pt}
\setlength{\abovedisplayskip}{4pt} \setlength{\abovedisplayshortskip}{4pt}
\begin{eqnarray}
&& A \equiv V(H,H) * \frac{S^H*(1-p)}{S^H*(1-p)+S^L*p} + V(H,L) * \frac{S^L*p}{S^H*(1-p)+S^L*p} - K. \label{imperfectsignaldynamic1}
\end{eqnarray}

When a high type does not signal, the payoff they get is given by the following expression:
\begin{eqnarray}
&& B \equiv V(H,H) * \frac{S^H*p}{S^H*p+S^L*(1-p)} + V(H,L) * \frac{S^L*(1-p)}{S^H*p+S^L*(1-p)}. \label{imperfectsignaldynamic2}
\end{eqnarray}

As such, in order for choosing to signal to be incentive compatible for high types the following relation must hold:
$(1-p)*A + p*B > p*A + (1-p)*B.$ Given that $p \in [0,1/2)$, this can be simplified to: $A > B$.


The inequality $A>B$ can be rewritten as:
\begin{eqnarray}
&& (V(H,H) - V(H,L))*(1-\frac{S^L*p}{S^H*(1-p)+S^L*p}-\frac{S^H*p}{S^H*p+S^L*(1-p)}) > K. \label{imperfectsignaldynamic6}
\end{eqnarray}

This inequality sets the upper bound on viable signal cost $K$. Since $p \in [0,1/2)$, and $V(H,H)>V(H,L)$, the left hand side term is positive if $S^L, S^H \neq 0$ which means there can exist a positive signaling cost $K$ under which high types are incentivized to signal. 

Likewise, when a low type does signal, the payoff they get is given by the following expression:
\begin{eqnarray}
&& C \equiv V(L,H) * \frac{S^H*(1-p)}{S^H*(1-p)+S^L*p} + V(L,L,) * \frac{S^L*p}{S^H*(1-p)+S^L*p} - K. \label{imperfectsignaldynamic7}
\end{eqnarray}

When a low type does not signal, the payoff they get is given by the following expression:
\begin{eqnarray}
&& D \equiv V(L,H) * \frac{S^H*p}{S^H*p+S^L*(1-p)} + V(L,L,) * \frac{S^L*(1-p)}{S^H*p+S^L*(1-p)}. \label{imperfectsignaldynamic8}
\end{eqnarray}

As such, in order for choosing to not signal to be incentive compatible for low types the following relation must hold: $(1-p)*C + p*D < p*C + (1-p)*D.$ Given that $p \in [0,1/2)$, this can be simplified to: $ C < D$. 
%

The inequality $C>D$ can be rewritten as:
\begin{eqnarray}
&& (V(L,H) - V(L,L))*(1-\frac{S^L*p}{S^H*(1-p)+S^L*p}-\frac{S^H*p}{S^H*p+S^L*(1-p)}) < K. \label{imperfectsignaldynamic12}
\end{eqnarray}

This inequality sets the lower bound for incentive compatible levels of $K$. Since $p \in [0,1/2)$, the term $1-\frac{S^L*p}{S^H*(1-p)+S^L*p}-\frac{S^H*p}{S^H*p+S^L*(1-p)}\in (0,1]$. In effect, an increase in $p$ decreases the lower bound on $K$. 

Combining inequalities [\ref{imperfectsignaldynamic6}] and [\ref{imperfectsignaldynamic12}] we get the following expression:
\begin{eqnarray}
&& 
\begin{split}
(V(H,H) - V(H,L))*(1-\frac{S^L*p}{S^H*(1-p)+S^L*p}-\frac{S^H*p}{S^H*p+S^L*(1-p)}) > K > \\
(V(L,H) - V(L,L))*(1-\frac{S^L*p}{S^H*(1-p)+S^L*p}-\frac{S^H*p}{S^H*p+S^L*(1-p)}).
\end{split}
\label{imperfectsignaldynamic13}
\end{eqnarray}

 \begin{figure}[h]
   \caption{Range of Viable Signal Costs (K) by Signal Accuracy (1-p) and Degree of Heterogeneity}
   \label{fig:ViableK}
    \includegraphics[width=\textwidth, height=.28\textheight]{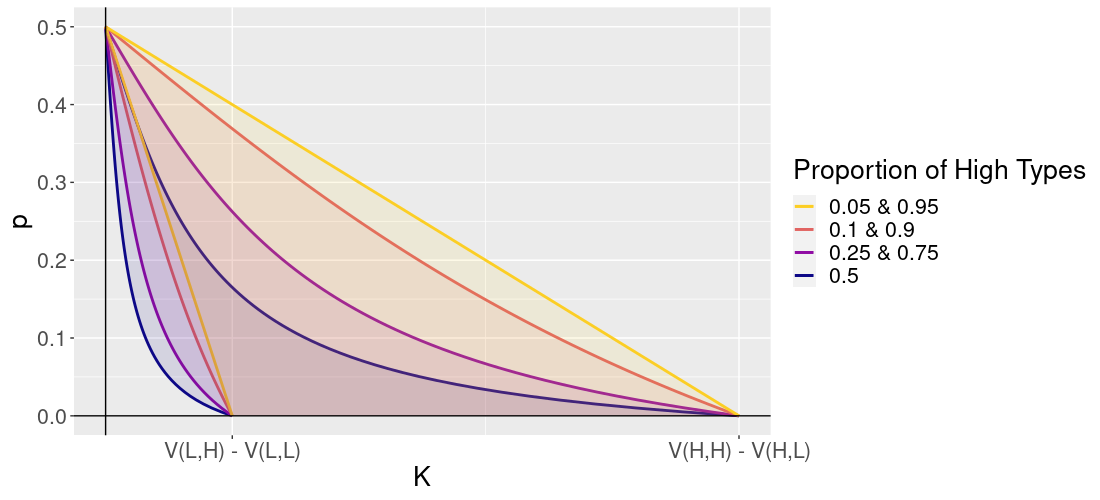}
    \end{figure}
 
 Figure \ref{fig:ViableK} shows how the accuracy of signaling ($1-p$) and the degree of heterogeneity within the entire population  (measured by $\min\{\frac{S^H}{S^H+S^L}, \frac{S^H}{S^H+S^L}\}$) affect viable levels of costly signaling ($K$) that give a separating equilibrium. Note that as the accuracy of signaling or heterogeneity decrease, the range of viable signal costs ($K$) decreases and approaches zero. This can be seen mathematically in equation [\ref{imperfectsignaldynamic13}] by taking the limit of the leftmost and rightmost terms as 1) $p$ approaches $1/2$, or as 2) $S^L$ or 3) $S^H$ approaches $0$. In all three cases, the limit of the terms is equal to $0$, thus collapsing the inequality. In each case, this is because the assortative benefit to signaling decreases as signaling accuracy or heterogeneity decreases. In the first case, when $p$ increases and approaches $1/2$, the decision to signal becomes less and less precise. As result, the assortativity within populations decreases. When signaling is imperfect ($p>0$) and heterogeneity decreases, the difference in type distribution between the signalers and non-signalers decreases. As $S^H$ or $S^L$ approaches $0$ (i.e., heterogeneity goes to $0$), the type distribution of signalers and that of non-signalers converge.
 
\begin{figure}[h]
   \caption{Range of Viable Signal Costs (K) and Equilibrium Type by Degree of Heterogeneity}
   \label{fig:Homogeneity}
    \includegraphics[width=\textwidth, height=.28\textheight]{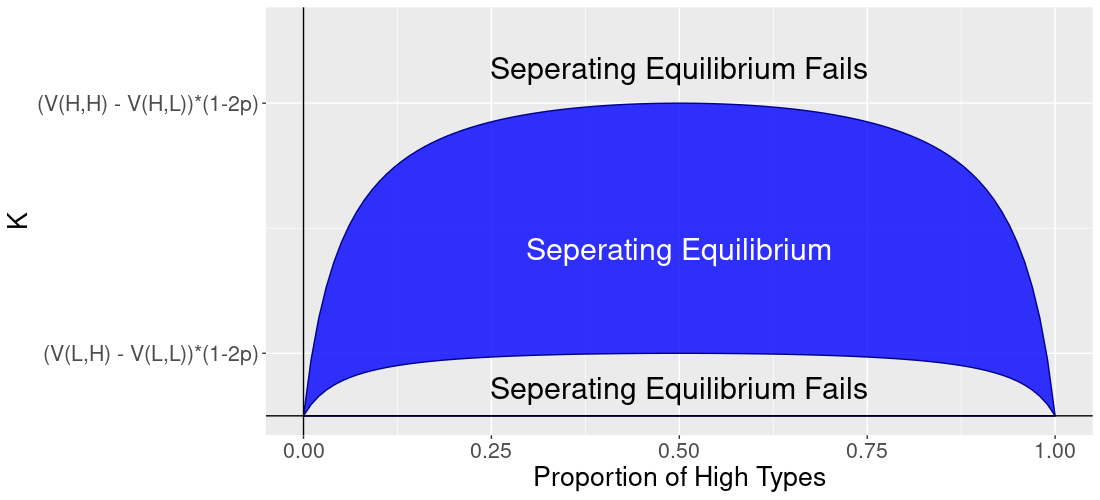}
    \end{figure}

 The assortative benefit of signaling when signaling is imperfect is maximized when the entire population is most heterogeneous. This can be seen in Figure \ref{fig:Homogeneity}. This graph shows the types of equilibria reached under different values of $K$ and degrees of heterogeneity with a fixed $p \in (0,1/2)$. This graph shows that as a population evolves from almost all low types to almost all high types with a fixed signal cost $K$, the type of equilibrium in the population can change, in some cases several times. First, suppose $K < (V(L,H) - V(L,L))*(1-2p)$. When the degree of heterogeneity is sufficiently low, the separating equilibrium inequality fails because the assortative benefit is not large enough for the high types to signal (as discussed in the previous paragraph). When the degree of heterogeneity is sufficiently high, the separating equilibrium inequality fails again  because it is no longer incentive compatible for the low types to not signal. Consequently, there exists two "Goldilocks" zones over which a separating equilibrium can exist. Second, suppose $(V(H,H) - V(H,L))*(1-2p) > K > (V(L,H) - V(L,L))*(1-2p)$. In this case,  there is no region where the low types would choose to signal and just one interval over which the high types would want to signal. Third, suppose $K > (V(H,H) - V(H,L))*(1-2p)$. In this case, the signaling cost is too large for a separating equilibrium to ever exist. 
 
Note that since $1-\frac{S^L*p}{S^H*(1-p)+S^L*p}-\frac{S^H*p}{S^H*p+S^L*(1-p)} \in (0,1]$ inequality [\ref{imperfectsignaldynamic13}] implies that in order for a separating equilibrium with signaling cost $K$ to exist, it must be the case that: \begin{equation}
     V(H,H)+V(L,L)>V(L,H)+V(H,L).
 \end{equation}
 This, is the same single crossing property that was derived when populations signal with perfect accuracy. \citep{Zahavi1975, Spence1973}.

\begin{figure}[p]
  \caption{Conflicts Between Imperfect Signaling and Non-Signaling Populations After a Period of Peace}
   \label{fig:IFight}
    \includegraphics[width=\textwidth, height=.28\textheight]{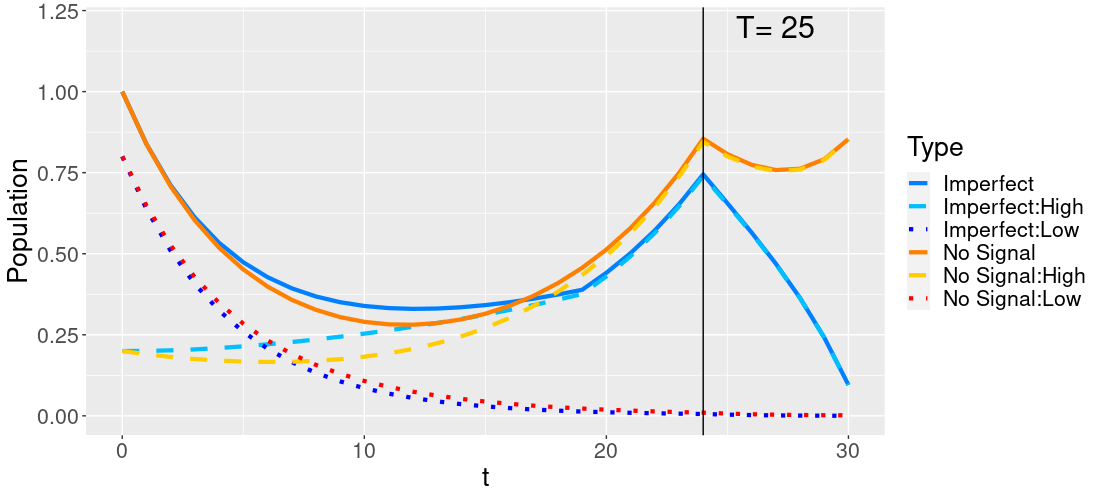}
    \includegraphics[width=\textwidth, height=.28\textheight]{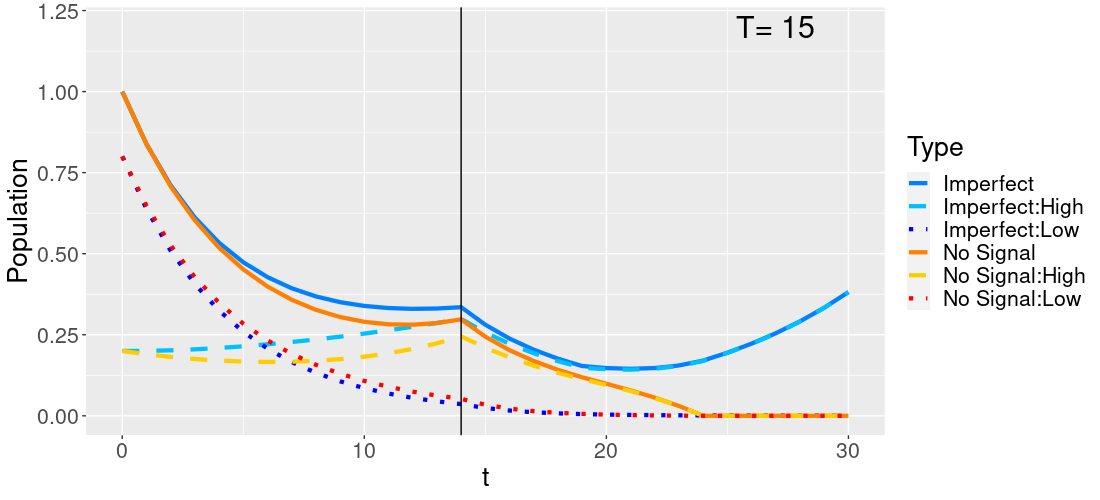}
    \includegraphics[width=\textwidth, height=.28\textheight]{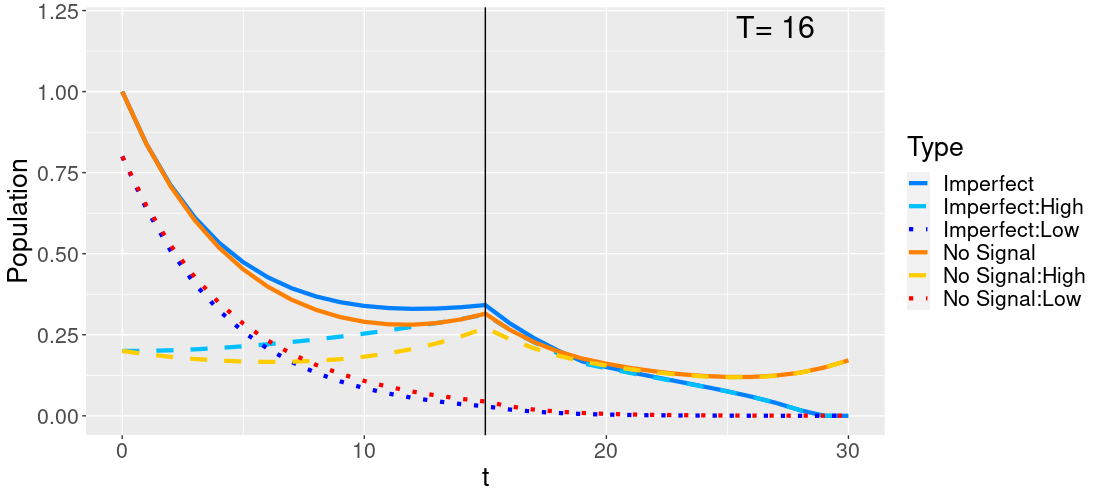}
 \begin{minipage}[c]{.2\textwidth}
    \textbf{Parameters:}
    \end{minipage}\hfill
    \begin{minipage}[c]{.2\textwidth}
    
    $N_0^H = S_0^H = .2$
    
    $N_0^L = S_0^L = .8$
    
    $K = .1$
    
    $\beta = .2$

    $p = .05$
    \end{minipage}\hfill
  \begin{minipage}[c]{.3\textwidth}
  \begin{tabular}{cc|c|c|}
      & \multicolumn{1}{c}{} & \multicolumn{1}{c}{H}  & \multicolumn{1}{c}{L} \\\cline{3-4}
      \multirow{2}*{}  & H & $1.15$ & $0.90$ \\\cline{3-4}
      & L & $0.85$ & $0.80$ \\\cline{3-4}
    \end{tabular}
    \end{minipage}\hfill
    \begin{minipage}[c]{.3\textwidth}
    The vertical lines indicate the last period before conflicts begin $(T-1)$.
    \end{minipage}
    \end{figure}

As in the case of perfect signaling, imperfect signaling can improve short run population growth through imperfect assortative matching resulting in faster evolution of the type makeup of a population. Unlike the case of perfect signaling, once the population is nearly entirely high types, the population no longer signals by choice, only signaling by mistake with probability $p$. However, this $p*K$ cost of signaling means that the non-signaling population will always, in the long run, attain a higher reproductive rate and larger population, just like in the case of perfect signaling. In examining how imperfect signaling might impact competition, Figure \ref{fig:IFight} provides three examples of outcomes when conflicts occur in our model, using the same initial parameters used in figure \ref{fig:Fight}, but here with a 5\% chance of making a mistake when signaling, that is, $p =0.05$. As can be seen, the margins between the populations in figure \ref{fig:IFight} become slimmer than in figure \ref{fig:Fight}, however, the same patterns remain. The imperfect signaling population may gain an advantage initially and if competition occurs, can defeat the non-signaling population. However, given enough time to evolve before conflict, the non-signaling population will always win.

\subsection{Concluding Remarks}

It has been well understood that costly signaling can provide a competitive advantage at the intra-group level, however, to our knowledge the effects of signaling has not been examined at the inter-group level of conflicts. We consider a model in which the same individuals (high types) that benefit from signaling would evolve even in the absence of signaling. Thus, when there are two populations, one with and one without signaling, it follows that they would both eventually evolve to be homogeneously high type. At that point, if the two populations were to engage in conflicts, the population without signaling would have a competitive advantage because they do not need to pay the cost of signaling. However, we argue that there can be an advantage to signaling in inter-group conflicts: that the group with signaling evolves faster than the group without signaling. As a result, if conflicts occur, the prevailing population is largely a question of how long the populations evolve in isolation before they engage in conflicts. Our model predicts that societies that have shorter periods of isolation before conflicts, and more efficient weapon used in conflicts may favor the rise of signaling norms.

\bibliographystyle{abbrvnat}
\bibliography{bib.bib}

\begin{thebibliography}{34}
\providecommand{\natexlab}[1]{#1}
\providecommand{\url}[1]{\texttt{#1}}
\expandafter\ifx\csname urlstyle\endcsname\relax
  \providecommand{\doi}[1]{doi: #1}\else
  \providecommand{\doi}{doi: \begingroup \urlstyle{rm}\Url}\fi

\bibitem[Bornstein(2003)]{Bornstein2003}
G.~Bornstein.
\newblock Intergroup conflict: Individual, group, and collective interests.
\newblock \emph{Personality and Social Psychology Review}, 7\penalty0
  (2):\penalty0 129--145, 2003.

\bibitem[Buss and Shackelford(1997)]{BussShackelford1997}
D.~M. Buss and T.~K. Shackelford.
\newblock Human aggression in evolutionary psychological perspective.
\newblock \emph{Clinical Psychological Review}, 17:\penalty0 605--619, 1997.

\bibitem[Clifton(2020)]{clifton2020brief}
E.~Clifton.
\newblock A brief review on the application of lanchester’s models of combat
  in nonhuman animals.
\newblock \emph{Ecological Psychology}, 32\penalty0 (4):\penalty0 181--191,
  2020.

\bibitem[Ferguson(2012)]{Ferguson2012}
B.~R. Ferguson.
\newblock Tribal walfare.
\newblock In G.~Martel, editor, \emph{The Encyclopedia of War}, pages
  2232--2244. Wiley-Blackwell, 2012.

\bibitem[Gat(2006)]{Gat2006}
A.~Gat.
\newblock \emph{War in human civilization}.
\newblock Oxford University Press, Oxford, 2006.

\bibitem[Getty(2006)]{Getty2006}
T.~Getty.
\newblock Sexually selected signals are not similar to sports handicaps.
\newblock \emph{Trends in ecology and evolution}, 21:\penalty0 83--88, 2006.

\bibitem[Gintis et~al.(2001)Gintis, Alden~Smith, and Bowles]{Gintisetal2001}
H.~Gintis, E.~Alden~Smith, and S.~Bowles.
\newblock Costly signaling and cooperation.
\newblock \emph{Journal of Theoretical Biology}, 213:\penalty0 103--119, 2001.

\bibitem[Grafen(1990)]{Grafen1990}
A.~Grafen.
\newblock Biological signals as handicaps.
\newblock \emph{Journal of Theoretical Biology}, 144:\penalty0 517--546, 1990.

\bibitem[Grose(2011)]{Grose2011}
J.~Grose.
\newblock Modelling and the fall and rise of the handicap principle.
\newblock \emph{Biology \& Philosophy}, 26:\penalty0 677--696, 2011.

\bibitem[Guilaine and Zammit(2004)]{GuilaineXammit2004}
J.~Guilaine and J.~Zammit.
\newblock \emph{The origins of war: Violence in prehistory}.
\newblock Blackwell, Oxford, 2004.

\bibitem[Hopkins(2014)]{Hopkins2014AEJMICRO}
E.~Hopkins.
\newblock Competitive altruism, mentalizing and signaling.
\newblock \emph{American Economic Journal: Microeconomics}, 2656\penalty0
  (4):\penalty0 272--292, 2014.

\bibitem[Johnson and MacKay(2015)]{JohnsonMackay2015EHB}
D.~D. Johnson and N.~J. MacKay.
\newblock Fight the power: Lanchester's laws of combat in human evolution.
\newblock \emph{Evolution and Human Behaviour}, 36:\penalty0 152--163, 2015.

\bibitem[Johnstone(1997)]{Johnstone1997}
R.~Johnstone.
\newblock The evolution of animal signals.
\newblock In J.~Krebs and N.~Davies, editors, \emph{Behavioral ecology, an
  evolutionary approach}, pages 465--485. Blackwell Scientific Publications,
  Oxford, 4 edition, 1997.

\bibitem[Keeley(1996)]{Keeley1996}
L.~Keeley.
\newblock \emph{War before civilization: The myth of the peaceful savage}.
\newblock Oxford University Press, Oxford, 1996.

\bibitem[Lanchester(1916)]{Lanchester1916}
F.~W. Lanchester.
\newblock \emph{Aircraft in warfare: The dawn of the fourth arm}.
\newblock Constable \& Co., London, 1916.

\bibitem[M.~Potts(2008)]{PottsHayden2008}
T.~H. M.~Potts.
\newblock \emph{Sex and war: How biology explains warfare and terrorism and
  offers a path to a safer world}.
\newblock Benbella Books, Dallas, TX, 2008.

\bibitem[Maynard~Smith and Harper(1995)]{MaynardSmithHarper1995}
J.~Maynard~Smith and D.~Harper.
\newblock Animal signals: models and terminology.
\newblock \emph{Journal of Theoretical Biology}, 177:\penalty0 305--311, 1995.

\bibitem[Maynard~Smith and Harper(2003)]{MaynardSmithHarper2003}
J.~Maynard~Smith and D.~Harper.
\newblock \emph{Animal Signals}.
\newblock Oxford University Press, Oxford, 2003.

\bibitem[Miller(2000)]{Miller2000}
G.~Miller.
\newblock \emph{The Mating Mind: How Sexual Choice Shaped the Evolution of
  Human Nature}.
\newblock Heinemann, London, 2000.

\bibitem[Nax and Rigos(2016)]{NaxRigos2016}
H.~H. Nax and A.~Rigos.
\newblock Assortativity evolving from social dilemmas.
\newblock \emph{Journal of Theoretical Biology}, 395:\penalty0 194--203, 2016.

\bibitem[Newton(2017)]{Newton2017IJGT}
J.~Newton.
\newblock The preferences of homo moralis are unstable under evolving
  assortativity.
\newblock \emph{International Journal of Game Theory}, 46:\penalty0 583--589,
  2017.

\bibitem[Przepiorka and Diekmann(2021)]{PrzepiorkaDiekmann2021}
W.~Przepiorka and A.~Diekmann.
\newblock Parochial cooperation and the emergence of signalling norms.
\newblock \emph{Philosophical Transactions of the Royal Society B: Biological
  Sciences}, 376:\penalty0 20200294, 2021.

\bibitem[Roberts(1998)]{Roberts1998}
G.~Roberts.
\newblock Competitive altruism: From reciprocity to the handicap principle.
\newblock \emph{Proceedings of the Royal Society B}, 265\penalty0
  (1395):\penalty0 427--431, 1998.

\bibitem[R.W.~Wrangham(1996)]{WranghamPeterson1996}
D.~P. R.W.~Wrangham.
\newblock \emph{Demonic males: Apes and the origins of human violence}.
\newblock Bloomsbury, London, 1996.

\bibitem[S.~LeBlanc(2003)]{LeBlancRegister2003}
K.~R. S.~LeBlanc.
\newblock \emph{Constant battles: The myth of the peaceful, noble savage}.
\newblock St. Martin's Press, New York, 2003.

\bibitem[Searcy and Nowicki(2005)]{SearcyNowicki2005}
W.~Searcy and S.~Nowicki.
\newblock \emph{The evolution of animal communication. Reliability and
  deception in signalling systems}.
\newblock Princeton University Press, Princeton, 2005.

\bibitem[Spence(1973)]{Spence1973}
A.~Spence.
\newblock Job market signaling.
\newblock \emph{Quarterly Journal of Economics}, 87:\penalty0 355--374, 1973.

\bibitem[Sz\'{a}mad\'{o}(2012)]{Szamado2012}
S.~Sz\'{a}mad\'{o}.
\newblock The rise and fall of the handicap principle: a commentary on the
  ``modelling and the fall and rise of the handicap principle.
\newblock \emph{Biology \& Philosophy}, 27:\penalty0 279--286, 2012.

\bibitem[Wilson et~al.(2002)Wilson, Britton, and Franks]{Wilsonetal2002}
M.~L. Wilson, N.~F. Britton, and N.~R. Franks.
\newblock Chimpanzees and the mathematics of battle.
\newblock \emph{Proceedings of the Royal Society B: Biological Sciences},
  269:\penalty0 1107--1112, 2002.

\bibitem[Wu(2016)]{wu2016}
J.~Wu.
\newblock Evolving assortativity and social conventions.
\newblock \emph{Economics Bulletin}, 36:\penalty0 936--941, 2016.

\bibitem[Wu(2018)]{wu2018}
J.~Wu.
\newblock Entitlement to assort: Democracy, compromise culture and economic
  stability.
\newblock \emph{Economics Letters}, 163:\penalty0 146--148, 2018.

\bibitem[Wu(2023)]{wu2023wp}
J.~Wu.
\newblock Institution, assortative matching and cultural evolution.
\newblock \emph{Working Paper}, 2023.

\bibitem[Zahavi(1975)]{Zahavi1975}
A.~Zahavi.
\newblock Mate selection: a selection for a handicap.
\newblock \emph{Journal of Theoretical Biology}, 53:\penalty0 205--214, 1975.

\bibitem[Zahavi and Zahavi(1997)]{ZahaviZahavi1997}
A.~Zahavi and A.~Zahavi.
\newblock \emph{The handicap principle}.
\newblock Oxford University Press, Oxford, 1997.

\end{thebibliography}

\end{document}